# Design of Energy Scavengers Mounted on Rotating Shafts #

S. M. Shahruz * and V. Sundararajan +

* Berkeley Engineering Research Institute, P. O. Box 9984, Berkeley, CA 94709

+ Department of Mechanical Engineering, University of California, Riverside, CA 92521

## Abstract

In this paper, a novel energy scavenger is proposed. The scavenger consists of a cantilever beam on which piezoelectric films and a mass are mounted. The mass at the tip of the beam is known as the proof mass and the device is called either an energy scavenger or a beam-mass system. The beam-mass system is mounted on a rotating shaft, where the axis of the shaft is horizontal. A single-degree-of-freedom (SDOF) mathematical model is derived for the scavenger and its properties are carefully examined. From the model, it becomes clear that the rotation of the shaft and gravity cause both parametric excitations and exogenous forces which make the beam-mass system vibrate. Guidelines are provided as how to choose the scavenger parameters in order to have it resonate. Examples are given to illustrate the performance of the proposed scavenger.

## 1. Introduction

In recent years, researchers have been designing and studying devices that convert the energy of vibration sources into electricity to power micro-electronic devices; see, e.g., Refs. [1-7] and references therein. Such devices are called energy scavengers (harvesters). Vibration sources the energy of which are scavenged are typically buildings, bridges, cars, trains, aircraft,

# This paper has been submitted to a journal for publication.

ships, manufacturing tools, etc. Transduction principles based on which the energy of vibration sources is converted into electricity are: electrostatics, electromagnetism, and piezoelectricity.

One type of energy scavenger designed based on the piezoelectricity principle is shown in Fig. 1. This device consists of a cantilever beam on which piezoelectric films and a mass are mounted. The mass at the tip of the cantilever beam is known as the proof mass and the device is called either an energy scavenger or a beam-mass system. When a scavenger is mounted on a vibration source, say a panel, the cantilever beam would vibrate. Due to the vibration of the beam, the piezoelectric films generate electric charge that can be converted into electric power.

By surveying the literature on the design of energy scavengers, it is concluded that beam-mass systems of most scavengers are modeled by an equation such as

$$m\,\ddot{y}(t) + c\,\dot{y}(t) + k\,y(t) = F(t)\,, \qquad y(0) = 0\,, \qquad \dot{y}(0) = 0 \tag{1}$$

for all $t \geq 0$, where $m$, $c$, and $k$ are, respectively, the effective mass, damping, and stiffness of the beam-mass system; $y(t) \in \mathbb{R}$ is the transversal displacement of the tip of the beam due to the scalar-valued transversal force $t \mapsto F(t)$.

For efficient energy scavenging, the beam-mass system should vibrate persistently with large amplitudes; see, e.g., Refs. [4, 7] and [8, p. 20]. In order to have large $y(\cdot)$, the peak-power frequency of the input force should match the resonant (fundamental) frequency of the beam-mass system. In this case, the beam-mass system resonates.

There are other approaches to the design of mechanical resonators; for instance, by mounting a beam-mass system vertically on a vibration source. In this case, the beam is under an axial force. It can be shown that the transversal displacement of the tip of the beam satisfies an equation such as (see, e.g., Ref. [9, pp. 77-81])

$$m\,\ddot{y}(t) + c\,\dot{y}(t) + [k - F_p(t)]\,y(t) = 0\,, \qquad y(0) \neq 0\,, \qquad \dot{y}(0) = 0 \tag{2}$$

for all $t \geq 0$, where $F_p(t)$ is due to the vibration source. The right-hand side of Eq. (2) is zero, while the excitation is in one of the system parameters. System (2) is said to be parametrically excited. By appropriate choices of $k$ and $t \mapsto F_p(t)$, it is possible to have system (2) resonate with large amplitudes. For instance, when $t \mapsto F_p(t)$ is a harmonic excitation, system (2) is known as the Mathieu system. It is well known that such a system can be destabilized by appropriate choices of $k$ and the amplitude of $F_p(\cdot)$, thereby having large amplitudes; see, e.g., Refs. [9, 10].

It is also possible to have a resonator where the transversal displacement of the tip of its beam satisfies

$$m\ \ddot{y}(t) + c\ \dot{y}(t) + [k - F_p(t)]\ y(t) = F(t)\,, \qquad y(0) = 0\,, \qquad \dot{y}(0) = 0 \tag{3}$$

for all $t \geq 0$ . In this case, the system is under the parametric excitation $F_p(\cdot)$ as well as the exogenous force $F(\cdot)$ . The energy scavenger to be designed in this paper is represented by an equation such as Eq. (3).

Regardless of how a beam-mass system is represented, either by Eq. (1) or (2) or (3), it is possible to alter the effective stiffness of such a system by using a magnetic proof mass and magnets that are placed in its vicinity. It is shown in Ref. [7] that when magnets are used, the beam-mass system in Eq. (1) can be represented by

$$m\ \ddot{y}(t) + c\ \dot{y}(t) + k\ y(t) - F_a(y(t)) = F(t)\,, \qquad y(0) = 0\,, \qquad \dot{y}(0) = 0 \tag{4}$$

for all $t \geq 0$ , where the nonlinear function $F_a(\cdot)$ represents the attractive force applied to the beam-mass system by the magnets. It is shown in Ref. [7] that by appropriate choice of $F_a(\cdot)$ , it is possible to arbitrarily tune the effective stiffness of the beam-mass system to lower values. In this case, the displacement of the tip of the beam is amplified and its resonant frequency is lowered. Larger amplitudes of vibration of beam-mass systems and the ability to tune the resonant frequencies of such systems arbitrarily are necessary for efficient energy scavenging from vibration sources.

In this paper, the goal is to design a novel energy scavenger which is mounted on rotating shafts. The rotation of the shaft and gravity would cause vibration in the scavenger. The organization of the paper is as follows. In Section 2, the proposed energy scavenger is described and a mathematical model representing its dynamics is derived. It turns out that the scavenger model is an equation such as that in Eq. (3). In Section 3, properties of the model that can be used in designing efficient energy scavengers are unveiled. In Section 4, a few examples are given to illustrate the performance of the scavenger.

## 2. Scavenger and its Mathematical Model

The energy scavenger proposed in this paper consists of a beam-mass system which is mounted on a rotating shaft; see, Fig. 2. The axis of rotation of the shaft is horizontal. The shaft rotates at different angular velocities and its angular displacement is the time function $t \mapsto \theta(t)$ .

The angular velocity of the shaft is usually constant during its operation. It will become clear later in the paper that the rotation of the shaft and gravity cause parametric excitations and exogenous forces applied to the beam-mass system.

A schematic of the beam-mass system mounted on the shaft is shown in Fig. 3. The length, width, and thickness of the beam are denoted by $l$, $w$, and $h$, respectively. The mass density and the modulus of elasticity of the beam are denoted by $\rho$ and $E$, respectively. The proof mass at the tip of the beam is assumed to be a point mass of mass $M$. The radius of the shaft is $R$.

Due to the rotation of the shaft, the beam vibrates transversally. The transversal displacement of the beam at an $x \in [0, l]$ and a $t \geq 0$ is denoted by $y(x, t) \in \mathbb{R}$.

With this setup, a mathematical model describing the dynamics of the beam-mass system is sought. There are numerous references that present models of a rotating beam with or without a point mass at its tip; see, e.g., Refs. [11, 12]. The available models are mostly complex and not suitable for the design of energy scavengers. What is desirable is a simple and mathematically tractable model; for instance, a single-degree-of-freedom (SDOF) model. Such a model will be obtained in the following by using the assumed-modes method; see, e.g., Refs. [13, pp. 308-312] and [14, pp. 242-245].

To apply the assumed-modes method, the transversal displacement of the beam is written as

$$y(x, t) = \phi(x)\, q(t) \tag{5}$$

for all $x \in [0, l]$ and $t \geq 0$, where the real- and scalar-valued function $x \mapsto \phi(x)$ is a known trial (shape) function and the real- and scalar-valued function $t \mapsto q(t)$ is an unknown (generalized coordinate) function. The trial function is chosen to be the first mode of the free transversal vibration of the beam-mass system in Fig. 3, given by (see, Ref. [15, p. 187, Table 6.6]):

$$\begin{aligned}\phi(x) = {} & \sin(\lambda(\alpha)(x/l)) - \sin\mathrm{h}(\lambda(\alpha)(x/l)) \\ & - \frac{\sin\lambda(\alpha) + \sin\mathrm{h}\lambda(\alpha)}{\cos\lambda(\alpha) + \cos\mathrm{h}\lambda(\alpha)}\,[\cos(\lambda(\alpha)(x/l)) - \cos\mathrm{h}(\lambda(\alpha)(x/l))]\end{aligned} \tag{6}$$

for all $x \in [0, l]$, where $\lambda$ depends on the ratio

$$\alpha := \frac{M}{m_b} \tag{7}$$

where $M$ is the mass of the proof mass and

$$m_b := \rho whl \tag{8}$$

is the mass of the beam. The dependence of $\lambda$ on $\alpha$ is given in Ref. [15, p. 188, Table 6.7(a)].

Using the first two terms in the series expansions of $\sin(\lambda(\alpha)(x/l))$, $\sinh(\lambda(\alpha)(x/l))$, $\cos(\lambda(\alpha)(x/l))$, and $\cosh(\lambda(\alpha)(x/l))$ (see, e.g., Ref. [16]) in Eq. (6), the trial function is approximated by

$$\phi(x) = a(\alpha)\left(\frac{x}{l}\right)^2 - b(\alpha)\left(\frac{x}{l}\right)^3 \tag{9}$$

for all $x \in [0, l]$, where

$$a(\alpha) = \frac{\sin\lambda(\alpha) + \sinh\lambda(\alpha)}{\cos\lambda(\alpha) + \cosh\lambda(\alpha)}\,\lambda^2(\alpha), \qquad b(\alpha) = \frac{1}{3}\,\lambda^3(\alpha). \tag{10}$$

Using the value of $\alpha$ and the corresponding $\lambda(\alpha)$ in Ref. [15, p. 188, Table 6.7(a)], $a(\alpha)$ and $b(\alpha)$ can be computed for all $\alpha \geq 0$. It can be easily verified that for any $\alpha \geq 0$, the functions in Eqs. (6) and (9) are very close to each other over the interval $[0, l]$ by plotting their graphs. In the following, for the sake of brevity, the dependence of $\lambda$, $a$, $b$, $x \mapsto \phi(x)$, and other functions on $\alpha$ is not stated explicitly everywhere.

In developing the SDOF model of the beam-mass system mounted on the rotating shaft, the kinetic and potential energies of the system should be available. These energies are obtained by using results from Refs. [11, 12].

The kinetic energy of the beam-mass system can be written as

$$T(t) = \frac{1}{2}\, m \int_0^l [(R + x)^2\, \dot\theta^2(t) + 2(R + x)\, \dot\theta(t)\, y_t(x, t) + y_t^2(x, t) + \dot\theta^2(t)\, y^2(x, t)]\, dx +$$

$$\frac{1}{2}\, M(R + l)^2\, \dot\theta^2(t) + M(R + l)\, \dot\theta(t)\, y_t(l, t) + \frac{1}{2}\, M\, y_t^2(l, t) + \frac{1}{2}\, M\, \dot\theta^2(t)\, y^2(l, t) \tag{11}$$

for all $t \geq 0$, where $m := \rho wh$ is the mass per unit length of the beam.

To obtain the potential energy of the beam-mass system, three potential energies should be considered. The first one corresponds to the elastic energy of the beam and is given by

$$V_e(t) = \frac{1}{2} \int_0^l EI \; y_{xx}^2(x,\, t)\; dx \tag{12}$$

for all $t \geq 0$, where $I = wh^3/12$ is the second moment of area of the beam cross section.

The second potential energy corresponds to the centrifugal forces and is given by

$$V_c(t) = \frac{1}{2} \int_0^l [mR\; \dot{\theta}^2(t)\; (l - x) + \frac{1}{2}\; m\dot{\theta}^2(t)\; (l^2 - x^2) + M(R + l)\; \dot{\theta}^2(t)]\; y_x^2(x,\, t)\; dx \tag{13}$$

for all $t \geq 0$.

The third potential energy is due to the gravity and is given by

$$V_g(t) = mgl \left( R + \frac{l}{2} \right) \sin\theta(t) - \frac{1}{2} \int_0^l mg \sin\theta(t)\; (l - x)\; y_x^2(x,\, t)\; dx\; +$$

$$\int_0^l mg \cos\theta(t)\; y(x,\, t)\; dx + Mg(R + l) \sin\theta(t) + Mg \cos\theta(t)\; y(l,\, t) \tag{14}$$

for all $t \geq 0$, where $g = 9.8\; m/s^2$ is the gravitational constant.

The potential energy of the beam-mass system is

$$V(t) = V_e(t) + V_c(t) + V_g(t) \tag{15}$$

for all $t \geq 0$.

Substituting Eq. (5), where $x \mapsto \phi(x)$ is that in Eq. (9), into Eqs. (11) and (15) and performing the necessary integrations, the kinetic and potential energies of the beam-mass system can be written, respectively, as

$$T(t) = \frac{1}{2} \left( \frac{m_b}{l} \right) \dot{\theta}^2(t) \int_0^l (R + x)^2\; dx$$

$$+\; m_b l\; A_7\; \dot{\theta}(t)\; \dot{q}(t) + \frac{1}{2}\; m_b\; A_1\; \dot{q}^2(t) + \frac{1}{2}\; m_b\; A_1\; \dot{\theta}^2(t)\; q^2(t)\; +$$

$$\frac{1}{2}\; M(R + l)^2\; \dot{\theta}^2(t) + M(R + l)\; A_2\; \dot{\theta}(t)\; \dot{q}(t) + \frac{1}{2}\; M\; A_2^2\; \dot{q}^2(t) + \frac{1}{2}\; M\; A_2^2\; \dot{\theta}^2(t)\; q^2(t) \tag{16}$$

$$V(t) = \frac{1}{2}\left(\frac{Ewh^3}{3l^3}\right)A_3\ q^2(t) +$$

$$\frac{1}{2}\ m_b\left(\frac{R}{l}\right)A_4\ \dot{\theta}^2(t)\ q^2(t) + \frac{1}{4}\ m_b\ A_5\ \dot{\theta}^2(t)\ q^2(t) + \frac{1}{2}\ M\left(\frac{R}{l}+1\right)A_6\ \dot{\theta}^2(t)\ q^2(t) +$$

$$m_b g\left(R + \frac{l}{2}\right)\sin\theta(t) - \frac{1}{2}\left(\frac{m_b}{l}\right)g\ A_4 \sin\theta(t)\ q^2(t) + m_b g\ A_8 \cos\theta(t)\ q(t) +$$

$$Mg(R+l)\sin\theta(t) + Mg\ A_2\cos\theta(t)\ q(t) \tag{17}$$

for all $t \geq 0$ , where $m_b$ is that in Eq. (8) and the dimensionless coefficients $A_1,\ A_2,\ \ldots,\ A_8$ are given by

$$A_1 := \frac{a^2}{5} - \frac{ab}{3} + \frac{b^2}{7} \tag{18a}$$

$$A_2 := a - b \tag{18b}$$

$$A_3 := a^2 - 3ab + 3b^2 \tag{18c}$$

$$A_4 := \frac{a^2}{3} - \frac{3ab}{5} + \frac{3b^2}{10} \tag{18d}$$

$$A_5 := \frac{8a^2}{15} - ab + \frac{18b^2}{35} \tag{18e}$$

$$A_6 := \frac{4a^2}{3} - 3ab + \frac{9b^2}{5} \tag{18f}$$

$$A_7 := \left(\frac{R}{l}\right)\left(\frac{a}{3} - \frac{b}{4}\right) + \left(\frac{a}{4} - \frac{b}{5}\right) \tag{18g}$$

$$A_8 := \frac{a}{3} - \frac{b}{4}\,. \tag{18h}$$

In deriving coefficients $A_1,\ A_2,\ \ldots,\ A_8$ , mathematical operations, mostly integration, were used. In Appendix A, it is explained how these coefficients are obtained.

Using the values of $\alpha$ and the corresponding $\lambda(\alpha)$ in Ref. [15, p. 188, Table 6.7(a)], $a$ and $b$ in Eq. (10) can be computed for all $\alpha \geq 0$ . Having these quantities computed, it can be verified numerically that $A_1,\ A_2,\ \ldots,\ A_8$ are non-negative.

The Lagrangian of the beam-mass system mounted on the rotating shaft is $L(t) = T(t) - V(t)$ for all $t \geq 0$, where $T(\cdot)$ and $V(\cdot)$ are given by Eqs. (16) and (17), respectively. Having the Lagrangian and using Lagrange's equation, while adding a viscous damping term, the equation of motion of the system is obtained as

$$m_e\, \ddot{q}(t) + c\, \dot{q}(t) + [K_e + K_c\, \dot{\theta}^2(t) - K_g\, g \sin\theta(t)]\, q(t) = -\, F_a\, \ddot{\theta}(t) - F_g\, g \cos\theta(t)$$

$$q(0) = 0\,, \qquad \dot{q}(0) = 0 \tag{19}$$

for all $t \geq 0$, where $c > 0$ is the damping coefficient and

$$m_e = m_b\, A_1 + M\, A_2^2 \tag{20a}$$

$$K_e = \left(\frac{Ewh^3}{3l^3}\right) A_3 \tag{20b}$$

$$K_c = m_b \left[\left(\frac{R}{l}\right) A_4 + \frac{1}{2}\, A_5 - A_1\right] + M \left[\left(\frac{R}{l} + 1\right) A_6 - A_2^2\right] \tag{20c}$$

$$K_g = \left(\frac{m_b}{l}\right) A_4 \tag{20d}$$

$$F_a = l \left[m_b\, A_7 + M \left(\frac{R}{l} + 1\right) A_2\right] \tag{20e}$$

$$F_g = m_b\, A_8 + M\, A_2\,. \tag{20f}$$

All constant coefficients in Eq. (19) are known. The time-varying coefficients and the exogenous input forces, respectively, on the left- and right-hand sides of Eq. (19) are known when $t \mapsto \theta(t)$ is known. It is clear that Eq. (19) has a form similar to that of Eq. (3). That is, system (19) is parametrically excited via the time-varying coefficients of $q(\cdot)$ and is under exogenous forces as well.

The rotation of the shaft has the indispensable role of causing the parametric excitations that appear as the coefficients of $q(\cdot)$ in Eq. (19). Only when the shaft rotates, does the gravity have a role in causing the parametric excitation $-\, K_g\, g \sin\theta(\cdot)$. Both rotation and gravity have roles in causing the exogenous forces, where the gravity plays a more important role.

If $\theta(t) = \Omega\, t$ for all $t \geq 0$, i.e., when the angular velocity of the shaft is constant, then $\dot{\theta}(t) = \Omega$ and $\ddot{\theta}(t) = 0$. In this case, the effect of the centrifugal force in Eq. (19) is the constant value $K_c\,\Omega^2$ , while the parametric excitation and the exogenous force are harmonic functions of time.

The representation of the beam-mass system mounted on the rotating shaft in Eq. (19) is a linear time-varying system. This equation can be solved for $t \mapsto q(t)$ . When this solution and $x \mapsto \phi(x)$ in Eq. (9) are substituted into Eq. (5), the transversal displacement of the beam, $y(x, t)$ , for all $x \in [0, l]$ and $t \geq 0$ is (approximately) obtained. For any $t \geq 0$ , the maximum of $y(x, t)$ is attained at $x^*(\alpha) = [2a(\alpha)/(3b(\alpha))]\, l$ . For any $\alpha \geq 0$ , the value of $\lambda(\alpha)$ is known from Ref. [15, p. 188, Table 6.7(a)]; so are $a(\alpha)$ and $b(\alpha)$ in Eq. (10). Having $a(\alpha)$ and $b(\alpha)$ , it can be verified numerically that $x^*(\alpha) > l$ for all $\alpha \geq 0$ . Thus, the absolute of the displacement of the tip of the beam, $|\, y(l, t)\,|$, is maximum for all $t \geq 0$ . This displacement is given by

$$y_l(t) := y(l, t) = \phi(l)\; q(t) = [a(\alpha) - b(\alpha)]\; q(t) = A_2(\alpha)\; q(t) \tag{21}$$

for all $t \geq 0$ , where $A_2(\alpha)$ is that in Eq. (18b).

## 3. Properties of the Scavenger Model

In this section, properties of the mathematical model of the energy scavenger in Eq. (19) are unveiled. Knowledge of these properties are guidelines for the design of efficient energy scavengers.

**3.1. Stiffening Effect of the Rotation:** In Eq. (19), the coefficient of $q(\cdot)$ represents the effective stiffness of the beam-mass system. If the shaft on which the beam is mounted were not rotating, then the stiffness of the beam-mass system was $K_e$ given in Eq. (20b). Coefficient $-\,K_g\; g\, \sin\theta(\cdot)$ of $q(\cdot)$ represents a parametric excitation of the beam-mass system due to the rotation and gravity. If, for instance, $\theta(t) = \Omega\, t$ for all $t \geq 0$ , then this parametric excitation is a harmonic function of time.

In Eq. (19), coefficient $K_c\, \dot{\theta}^2(\cdot)$ of $q(\cdot)$ is due to the rotation and plays an important role. In the following, it will be shown that $K_c\, \dot{\theta}^2(t)$ is non-negative for all $t \geq 0$ . This implies that the rotation stiffens the beam.

**Fact 3.1:** In Eq. (19), coefficient $K_c\, \dot{\theta}^2(t)$ of $q(\cdot)$ is non-negative for all $t \geq 0$ .

**Proof:** This fact is established by showing that $K_c$ in Eq. (20c) is positive. Substituting $A_1$, $A_2$, $A_4$, $A_5$, and $A_6$ from Eq. (18) into Eq. (20c), it follows that

$$K_c = m_b \left[ \frac{R}{l} \left( \frac{a^2}{3} - \frac{3ab}{5} + \frac{3b^2}{10} \right) + \frac{a^2}{15} - \frac{ab}{6} + \frac{4b^2}{35} \right] +$$

$$M \left[ \frac{R}{l} \left( \frac{4a^2}{3} - 3ab + \frac{9b^2}{5} \right) + \frac{a^2}{3} - ab + \frac{4b^2}{5} \right]. \tag{22}$$

By using the vector $v^T = [a \quad b] \in \mathbb{R}^{1 \times 2}$, Eq. (22) can be written as

$$K_c = m_b v^T \left( \frac{R}{l} \begin{bmatrix} 1/3 & -3/10 \\ -3/10 & 3/10 \end{bmatrix} + \begin{bmatrix} 1/15 & -1/12 \\ -1/12 & 4/35 \end{bmatrix} \right) v +$$

$$M v^T \left( \frac{R}{l} \begin{bmatrix} 4/3 & -3/2 \\ -3/2 & 9/5 \end{bmatrix} + \begin{bmatrix} 1/3 & -1/2 \\ -1/2 & 4/5 \end{bmatrix} \right) v \,. \tag{23}$$

It can be easily verified that all matrices in Eq. (23) are positive definite. Therefore, $K_c > 0$ . □

Fact 3.1 establishes the stiffening of the beam due to the rotation, which is intuitively obvious.

**3.2. Effects of Parameters on the Scavenger Dynamics:** In order to design efficient energy scavengers, it is necessary to determine the effects of the scavenger parameters on its dynamics. To do so, Eq. (19) is divided by $m_e$ . The result is

$$\ddot{q}(t) + \tilde{c}\, \dot{q}(t) + [\tilde{K}_e + \tilde{K}_c\, \dot{\theta}^2(t) - \tilde{K}_g\, g \sin\theta(t)]\, q(t) = - \tilde{F}_a\, \ddot{\theta}(t) - \tilde{F}_g\, g \cos\theta(t)$$

$$q(0) = 0 \,, \qquad \dot{q}(0) = 0 \tag{24}$$

for all $t \geq 0$ , where

$$\tilde{c} = \frac{c}{\rho w h l} \left( \frac{1}{A_1(\alpha) + \alpha\, A_2^2(\alpha)} \right) \tag{25a}$$

$$\tilde{K}_e = \frac{E h^2}{3 \rho l^4} \left( \frac{A_3(\alpha)}{A_1(\alpha) + \alpha\, A_2^2(\alpha)} \right) \tag{25b}$$

$$\tilde{K}_c = \frac{(R/l)\ A_4(\alpha) + A_5(\alpha)/2 - A_1(\alpha) + \alpha\ [(R/l+1)\ A_6(\alpha) - A_2^2(\alpha)]}{A_1(\alpha) + \alpha\ A_2^2(\alpha)} \tag{25c}$$

$$\tilde{K}_g = \frac{1}{l}\left(\frac{A_4(\alpha)}{A_1(\alpha) + \alpha\ A_2^2(\alpha)}\right) \tag{25d}$$

$$\tilde{F}_a = l\left(\frac{A_7(\alpha) + \alpha(R/l+1)\ A_2(\alpha)}{A_1(\alpha) + \alpha\ A_2^2(\alpha)}\right) \tag{25e}$$

$$\tilde{F}_g = \frac{A_8(\alpha) + \alpha\ A_2(\alpha)}{A_1(\alpha) + \alpha\ A_2^2(\alpha)} \tag{25f}$$

and $\alpha$ is that in Eq. (7).

Coefficients of system (24) depend on several parameters. Among them $l$, $R/l$, and $\alpha$ play more important roles in the scavenger dynamics. The roles of these parameters are studied in the following.

**3.2.1. Beam length $l$:** This parameter appears in Eqs. (25a), (25b), (25d), and (25e). From these equations, it is clear that:

**(i)** Coefficients $\tilde{c}$, $\tilde{K}_e$, and $\tilde{K}_g$ are monotonically decreasing functions of $l$.

**(ii)** Coefficient $\tilde{F}_a$ is a monotonically increasing function of $l$.

By increasing $l$, coefficient $\tilde{K}_e$ decreases sharply. This property is useful when it is desired to have a beam-mass system with a low resonant frequency. By increasing $l$, however, $\tilde{K}_g$ decreases and so does role of the parametric excitation.

It is remarked that shafts mostly operate at constant angular velocities. Therefore, $-\tilde{F}_a\ \ddot{\theta}(t) = 0$ for all $t \geq 0$, and a larger $\tilde{F}_a$ for a larger $l$ plays no role in the beam-mass system response.

**3.2.2. Ratio $R/l$:** The ratio of the shaft radius to the beam length appears in Eqs. (25c) and (25e). Since $A_2(\alpha)$, $A_4(\alpha)$, and $A_6(\alpha)$ are positive for all $\alpha \geq 0$, coefficients $\tilde{K}_c$ and $\tilde{F}_a$ are monotonically increasing functions of $R/l$.

For a constant angular velocity $\dot{\theta}(t) = \Omega$ for all $t \geq 0$, the resonant frequency of the beam-mass system is $(\tilde{K}_e + \tilde{K}_c\ \Omega^2)^{1/2}$. By lowering $R/l$, the contribution of $\tilde{K}_c\ \Omega^2$ to the resonant frequency decreases.

**3.2.3. Proof mass $M$ (equivalently $\alpha$):** All coefficients of system (24) depend on $\alpha$. Functions of $\alpha$ that appear on the right-hand side of Eq. (25) are designated by the following scalar-valued functions:

$$\psi_1(\alpha) := \frac{1}{A_1(\alpha) + \alpha\ A_2^2(\alpha)} \tag{26a}$$

$$\psi_2(\alpha) := \frac{A_3(\alpha)}{A_1(\alpha) + \alpha\ A_2^2(\alpha)} \tag{26b}$$

$$\psi_3(\alpha) := \frac{(R/l)\ A_4(\alpha) + A_5(\alpha)/2 - A_1(\alpha) + \alpha\ [(R/l + 1)\ A_6(\alpha) - A_2^2(\alpha)]}{A_1(\alpha) + \alpha\ A_2^2(\alpha)} \tag{26c}$$

$$\psi_4(\alpha) := \frac{A_4(\alpha)}{A_1(\alpha) + \alpha\ A_2^2(\alpha)} \tag{26d}$$

$$\psi_5(\alpha) := \frac{A_7(\alpha) + \alpha(R/l + 1)\ A_2(\alpha)}{A_1(\alpha) + \alpha\ A_2^2(\alpha)} \tag{26e}$$

$$\psi_6(\alpha) := \frac{A_8(\alpha) + \alpha\ A_2(\alpha)}{A_1(\alpha) + \alpha\ A_2^2(\alpha)}\ . \tag{26f}$$

The functions in Eq. (26) are evaluated by using the values of $\alpha$ and the corresponding values of $\lambda(\alpha)$ in Ref. [15, p. 188, Table 6.7(a)] and Eqs. (10) and (18). The graphs of these functions versus $\alpha$ are plotted in Figs. 4-9.

The following conclusions are drawn from Figs. 4-9:

**(i)** Coefficient $\tilde{c}$ increases as $\alpha$ increases, except for all $\alpha \in [0, 0.25]$, where it decreases. (See Fig. 4.)

**(ii)** Coefficient $\tilde{K}_e$ is a monotonically decreasing function of $\alpha$. (See Fig. 5.)

**(iii)** Coefficient $\tilde{K}_c$ is a monotonically decreasing function of $\alpha$ for all $R/l \geq 0.2$. Note the dependence of $\tilde{K}_c$ on $R/l$ in Fig. 6. Small values of $R/l$ are not of interest since they correspond to shaft radii much smaller than beam lengths.

**(iv)** Coefficient $\tilde{K}_g$ is a monotonically decreasing function of $\alpha$. (See Fig. 7.)

**(v)** Coefficient $\tilde{F}_a$ is a monotonically increasing function of $\alpha$. Note the dependence of $\tilde{F}_a$ on $R/l$ in Fig. 8.

**(vi)** Coefficient $\tilde{F}_g$ is a monotonically increasing function of $\alpha$. (See Fig. 9.)

The fact that $\tilde{K}_e$ and $\tilde{K}_c$ decrease as $\alpha$ increases can be used as a design guideline. As it was stated earlier for a constant angular velocity $\dot{\theta}(t) = \Omega$ for all $t \geq 0$, the resonant frequency of the beam-mass system is $(\tilde{K}_e + \tilde{K}_c\ \Omega^2)^{1/2}$. Therefore, by increasing the mass of the proof mass (equivalently $\alpha$), the resonant frequency decreases.

**3.3. Effect of $A_2(\alpha)$ on Beam Displacement:** From Eq. (24), it is evident that the displacement of the tip of the beam depends linearly on $\tilde{F}_g(\alpha)$. From Eq. (19), this displacement also depends linearly on $A_2(\alpha)$.

From Fig. 9, it is clear that $\tilde{F}_g(\alpha)$ increases monotonically as $\alpha$ increases. Therefore, it is important to know the dependence of $A_2(\alpha)$, and more importantly, the dependence of $A_2(\alpha)\,\tilde{F}_g(\alpha)$ on $\alpha$.

In Fig. 10, the graph of $A_2(\alpha)$ versus $\alpha$ is shown. Clearly, $A_2$ is a monotonically decreasing function of $\alpha$. In Fig. 11, the graph of $A_2(\alpha)\,\tilde{F}_g(\alpha)$ versus $\alpha$ is shown. From this figure, it is evident that $A_2(\alpha)\,\tilde{F}_g(\alpha)$ decreases monotonically from $1.53$ to $1$ as $\alpha$ increases. This decrease is small. The good news is that $A_2(\alpha)\,\tilde{F}_g(\alpha)$ remains larger than $1$ for all $\alpha \geq 0$. Therefore, increasing $\alpha$ will not reduce the displacement of the tip of the beam.

From the detailed analysis of this section, it is concluded that:

**(i)** By increasing the length $l$ of the beam, the resonant frequency of the beam-mass system in Eq. (24) decreases sharply via $\tilde{K}_e$.

**(ii)** When the angular velocity of the shaft is a constant value $\Omega$, by decreasing the ratio $R/l$ of the shaft radius to the beam length, the contribution of $\tilde{K}_c\,\Omega^2$ to the resonant frequency of the beam-mass system decreases.

**(iii)** By increasing the mass $M$ of the proof mass (equivalently $\alpha$), the the resonant frequency of the beam-mass system decreases via $\tilde{K}_e$. Furthermore, when the angular velocity of the shaft is a constant value $\Omega$, larger proof masses decrease the contribution of $\tilde{K}_c\,\Omega^2$ to the resonant frequency.

**(iv)** The parametric excitation $-\tilde{K}_g\,g\,\sin\theta(\cdot)$ and the exogenous input $-\tilde{F}_a\,\ddot{\theta}(\cdot)$ do not play major roles in the scavenger dynamics.

## 4. Examples

In this section, a few examples are given to illustrate the performance of the energy scavenger proposed in this paper.

Let the beam in Fig. 2 be made out of silver with the following properties:

$$\rho = 10500\ kg/m^3\ , \qquad E = 7.8 \text{ x } 10^{10}\ N/m^2 \tag{27a}$$

$$c = 0.01\ Ns/m\ . \tag{27b}$$

Let the beam dimensions and the proof mass be

$$l = 0.05\ m\ , \qquad w = 0.005\ m\ , \qquad l = 0.001\ m \tag{28a}$$

$$M = 0.013125\ kg\ . \tag{28b}$$

For the parameter values in Eq. (28), it follows that $\alpha = M/\rho whl = 5$ . Therefore, from Ref. [15, p. 180, Table 6.7(a)], $\lambda = 0.87$ .

The beam-mass system is mounted on the shaft of radius

$$R = 0.02\ m\ . \tag{29}$$

The shaft rotates, where its angular displacement is the function $t \mapsto \theta(t)$ .

**Example 4.1:** Let the angular velocity of the shaft be the constant value $\dot{\theta}(t) = 900\ RPM$ for all $t \geq 0$ . With the parameter values in Eqs. (27)-(29), system (24) was simulated. The time history of the displacement of the tip of the beam, $y_l(\cdot)$ , is shown in Fig. 12. This displacement is a harmonic function of time after a transient. The amplitude of this function is $0.1865\ mm$ . □

When the angular velocity of the shaft is a constant value $\Omega$ , the resonant frequency of system (24) is $(\tilde{K}_e + \tilde{K}_c\ \Omega^2)^{1/2}$ and the exogenous input to the system is $-\tilde{F}_g\ g \cos \Omega\ t$ for all $t \geq 0$ . If the excitation frequency matches the resonant frequency, i.e., if

$$\tilde{K}_e + \tilde{K}_c\ \Omega^2 = \Omega^2 \tag{30}$$

then the beam-mass system resonates. This is a desirable condition since when it holds the beam would vibrate with large amplitudes. It is possible to have Eq. (30) hold by appropriate choices of $l$ , $R/l$ , and $\alpha$ . In the next example, such a possibility is explored.

**Example 4.2:** Let the setup in Eqs. (27)-(29) hold, while the constant angular velocity $\Omega$ of the shaft is unknown. The angular velocity $\Omega$ is sought so that Eq. (30) would hold. It turns out that this equation holds for $\Omega = 4049\ RPM$ . For this angular velocity, system (24) was simulated. The time history of the displacement of the tip of the beam, $y_l(\cdot)$ , is shown in Fig. 13. This displacement is a harmonic function of time after a transient. The amplitude of this function is $5.9\ mm$ . Due to resonance, the amplitude is quite large.

It is remarked that for constant angular velocities between $900\ RPM$ and $4049\ RPM$, system (24) was simulated. It is reported that when the angular velocity was increased from one constant value to the next, the amplitude of the tip of the beam increased. Beyond $4049\ RPM$, the amplitude of the tip of the beam decreased. □

**Example 4.3:** In this example, the performance of the scavenger is studied when the angular velocity of the shaft starts from zero and settles at a constant value $\Omega$.

Let the setup in Eqs. (27)-(29) hold, while the angular displacement, velocity, and acceleration of the shaft are given, respectively, by

$$\theta(t) = \Omega t\,(1 - e^{-t}) \tag{31a}$$

$$\dot{\theta}(t) = \Omega\,(1 - e^{-t} + te^{-t}) \tag{31b}$$

$$\ddot{\theta}(t) = \Omega\,(2e^{-t} - te^{-t}) \tag{31c}$$

for all $t \geq 0$. From Eq. (31), it is clear that the angular velocity of the shaft starts from zero and after a transient settles at the constant value $\Omega$.

With the parameter values in Eqs. (27)-(29) and the angular displacement, velocity, and acceleration in Eq. (31), where $\Omega = 2700\ RPM$, system (24) was simulated. The time history of the displacement of the tip of the beam, $y_l(\cdot)$, is shown in Fig. 14. This displacement is a harmonic function of time after a transient. The amplitude of this function is $0.3189\ mm$. The transient vibration has large amplitude □

From these examples, it is clear that the an energy scavenger can be designed by mounting a beam-mass system on a rotating shaft. By appropriate choices of the parameters of the beam-mass system and the shaft, it is possible to have the system vibrate with large amplitudes.

## 5. Conclusions

In this paper, a novel energy scavenger was proposed. In general, energy scavengers are designed to convert the energy of vibration sources into electricity. A class of scavengers does such a conversion by using piezoelectric films. A scavenger of this type consists of a cantilever beam on which piezoelectric films and a mass are mounted. The mass at the tip of the beam is known as the proof mass and the device is called either an energy scavenger or a beam-mass system.

Energy scavengers are mostly mounted on sources of random vibration. However, in experimental study of scavengers or simulation of their mathematical models, harmonic input forces are commonly used. The scavenger designed in this paper, has an inherent source of harmonic force.

The proposed energy scavenger is a beam-mass system mounted on a rotating shaft, where the axis of the shaft is horizontal. A single-degree-of-freedom (SDOF) mathematical model for the scavenger was derived using the assumed-modes method. The model is a linear time-varying system, which is under both parametric excitations and exogenous forces. The rotation of the shaft and gravity are the sources of the parametric excitations and exogenous forces. The model of the scavenger was studies in detail to unravel its properties. Knowledge of such properties are guidelines for the design of efficient energy scavengers.

Several examples were given to illustrate the performance of the scavenger. In one example, it was shown that the scavenger can resonate with large amplitudes. These examples prove that efficient energy scavengers can be made by mounting beam-mass systems on rotating shafts and by tuning their parameters appropriately.

From the examples given in this paper, it became evident that the parametric excitation and the exogenous force due to the angular acceleration of the shaft do not play major roles in the scavenger dynamics. Work is underway to determine whether the parametric excitation can be used to destabilize the beam-mass system, thereby its amplitude of vibration could be amplified.

## References

[1] Anton, S. P., and Sodano, H. A., 2007, "A Review of Power Harvesting Using Piezoelectric Materials (2003-2006)," Smart Materials and Structures, **16** (3), pp. R1-R21.

[2] Clark, W. W., 2005, "Special Issue on Energy Harvesting," Journal of Intelligent Material Systems and Structures, **16** (10), p. 783.

[3] Roundy, S., Wright, P. K., and Rabaey, J., 2004, *Energy Scavenging for Wireless Sensor Networks: With Special Focus on Vibrations*, Kluwer Academic Publishers, Boston, MA.

[4] Shahruz, S. M., 2006, "Design of Mechanical Band-Pass Filters With Large Frequency Bands for Energy Scavenging," Mechatronics, **16** (9), pp. 523-531.

[5] Shahruz, S. M., 2008, "Design of Mechanical Band-Pass Filters for Energy Scavenging: Multi-Degree-of-Freedom Models," J. Vib. Control, **14** (5), pp. 753-768.

[6] Shahruz, S. M., 2008, "Performance of Mechanical Bandpass Filters Used in Energy Scavenging in the Presence of Fabrication Errors and Coupling," Journal of Vibration and Acoustics, **130** (5), Paper # 054505.

[7] Shahruz, S. M., 2008, "Increasing the Efficiency of Energy Scavengers by Magnets," Journal of Computational and Nonlinear Dynamics, **3** (4), Paper # 041001.

[8] Jeon, Y. B., Sood, R., Jeong, J.-H., and Kim, S.-G., 2005, "MEMS Power Generator With Transverse Mode Thin Film PZT," Sens. Actuators, A, **122** (1), pp. 16-22.

[9] Cartmell, M., 1990, *Introduction to Linear, Parametric and Nonlinear Vibrations*, Chapman and Hall, London, UK.

[10] Jordan, D. W., and Smith, P., 2007, *Nonlinear Ordinary Differential Equations: An Introduction for Scientists and Engineers*, 4th edition, Oxford University Press, Oxford, UK.

[11] Lee, H. P., 1994, "Effect of Gravity on the Stability of a Rotating Cantilever Beam in a Vertical Plane," Computers & Structures, **53** (2), pp. 351-355.

[12] Yigit, A., Scott, R. A., and Galip Ulsoy, A., 1988, "Flexural Motion of a Radially Rotating Beam Attached to a Rigid Body," Journal of Sound and Vibration, **121** (2), pp. 201-210.

[13] Meirovitch, L., 1967, *Analytical Methods in Vibration*, MacMillan, New York, NY.

[14] Shabana, A. A., 1991, *Theory of Vibration, Volume II: Discrete and Continuous Systems*, Springer-Verlag, New York, NY.

[15] Karnovsky, I. A., and Lebed, O. I., 2004, *Free Vibration of Beams and Frames*, McGraw-Hill, New York, NY.

[16] Gradshteyn, I. S., and Ryzhik, I. M., 2000, *Table of Integrals, Series and Products*, 6th edition, Academic Press, San Diego, CA.

## Appendix A

Coefficients $A_1, A_2, \ldots, A_8$ in Eq. (18) are derived from the following mathematical expressions:

$$A_1 \quad \text{is derived from} \quad \int_0^l \phi^2(x)\, dx = l\, A_1 \tag{A.1}$$

$$A_2 \quad \text{is derived from} \quad \phi(l) = A_2 \tag{A.2}$$

$$A_3 \quad \text{is derived from} \quad \int_0^l \phi_{xx}^2(x)\, dx = \frac{4\, A_3}{l^3} \tag{A.3}$$

$$A_4 \quad \text{is derived from} \quad \int_0^l (l - x)\, \phi_x^2(x)\, dx = A_4 \tag{A.4}$$

$$A_5 \quad \text{is derived from} \quad \int_0^l (l^2 - x^2)\, \phi_x^2(x)\, dx = l\, A_5 \tag{A.5}$$

$$A_6 \quad \text{is derived from} \quad \int_0^l \phi_x^2(x)\, dx = \frac{A_6}{l} \tag{A.6}$$

$$A_7 \quad \text{is derived from} \quad \int_0^l (R + x)\, \phi(x)\, dx = l^2\, A_7 \tag{A.7}$$

$$A_8 \quad \text{is derived from} \quad \int_0^l \phi(x)\, dx = l\, A_8 \tag{A.8}$$

where $x \mapsto \phi(x)$ is that in Eq. (9).

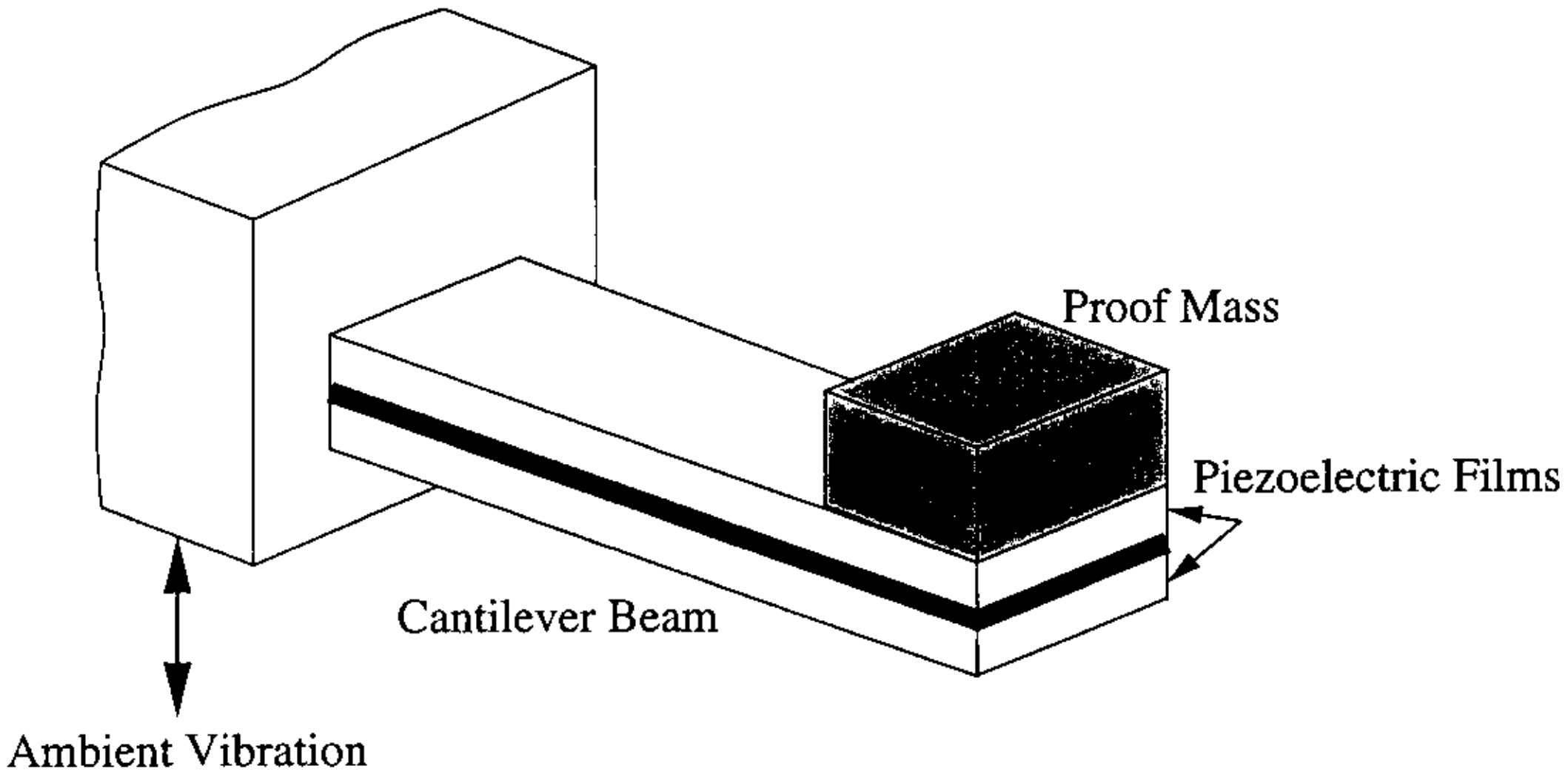

**Fig. 1:** A typical energy scavenger consists of a cantilever beam on which piezoelectric films and a mass, known as the proof mass, are mounted.

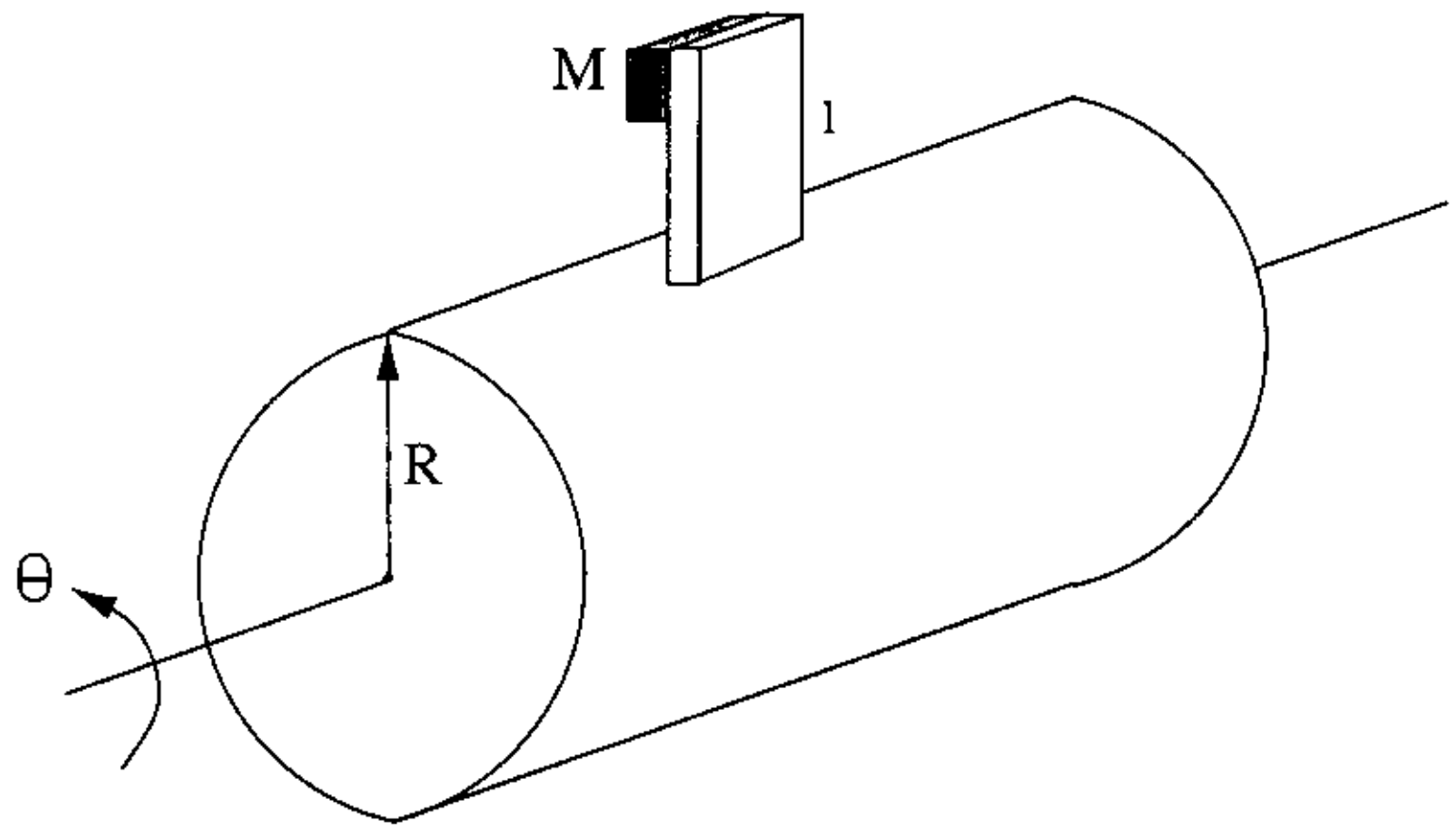

**Fig. 2:** A beam-mass system mounted on a rotating shaft. The axis of rotation of the shaft is horizontal. The shaft rotates at different angular velocities and its angular displacement is the time function $t \mapsto \theta(t)$ .

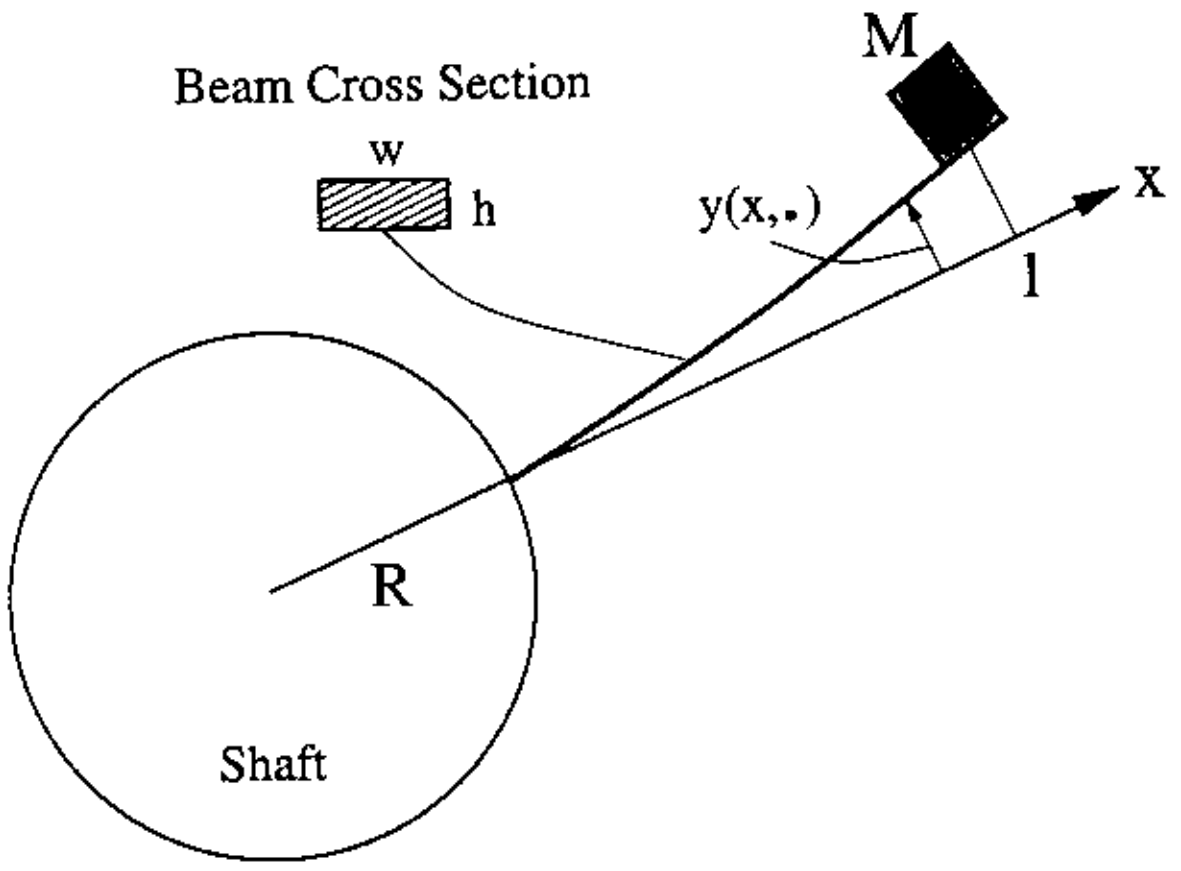

**Fig. 3:** A schematic of a beam with a proof mass at its tip. The beam-mass system is mounted on a rotating shaft. The transversal displacement of the beam at an $x \in [0, l]$ and a $t \geq 0$ is denoted by $y(x, t)$ .

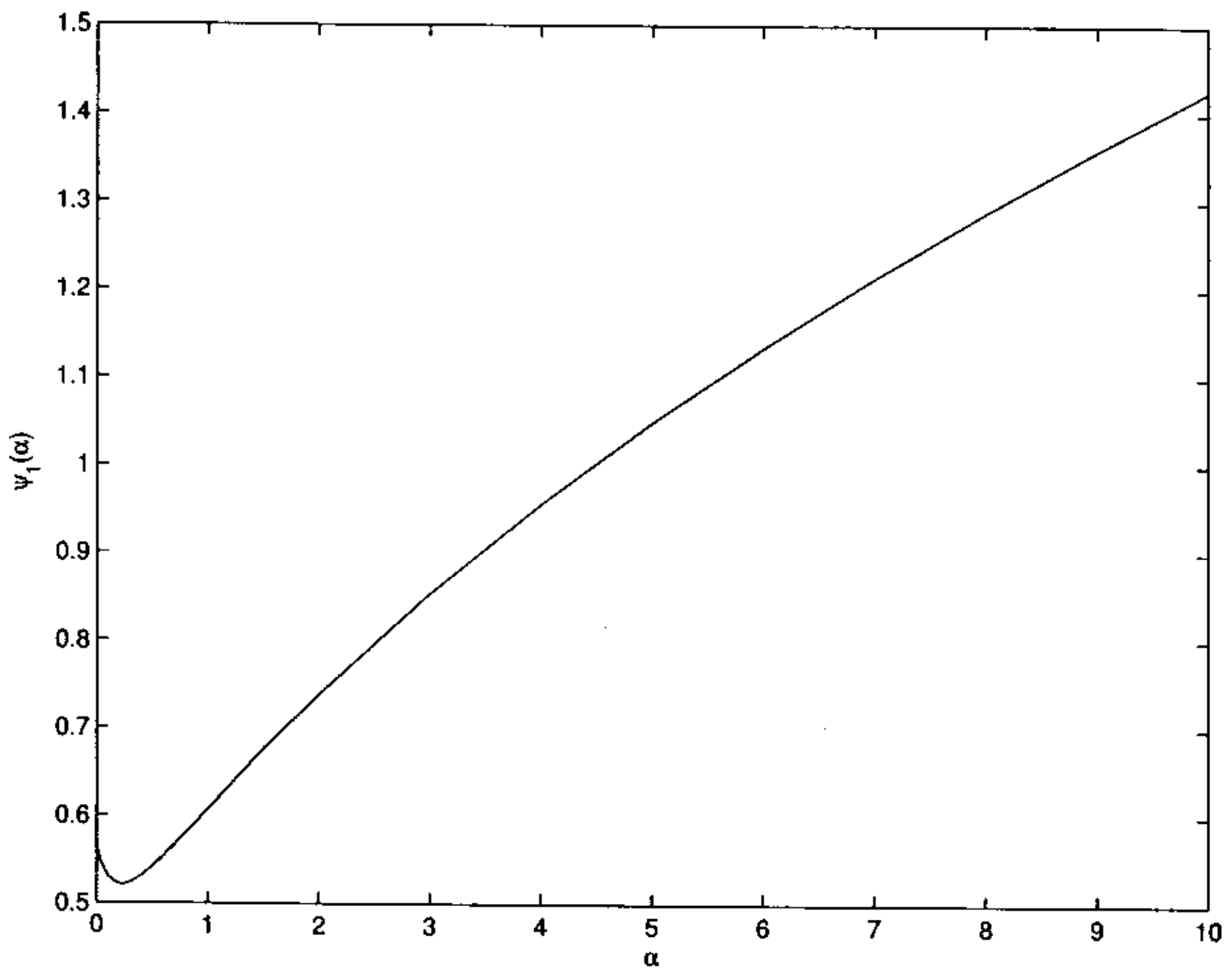

**Fig. 4:** Graph of function $\alpha \mapsto \psi_1(\alpha)$ in Eq. (26a).

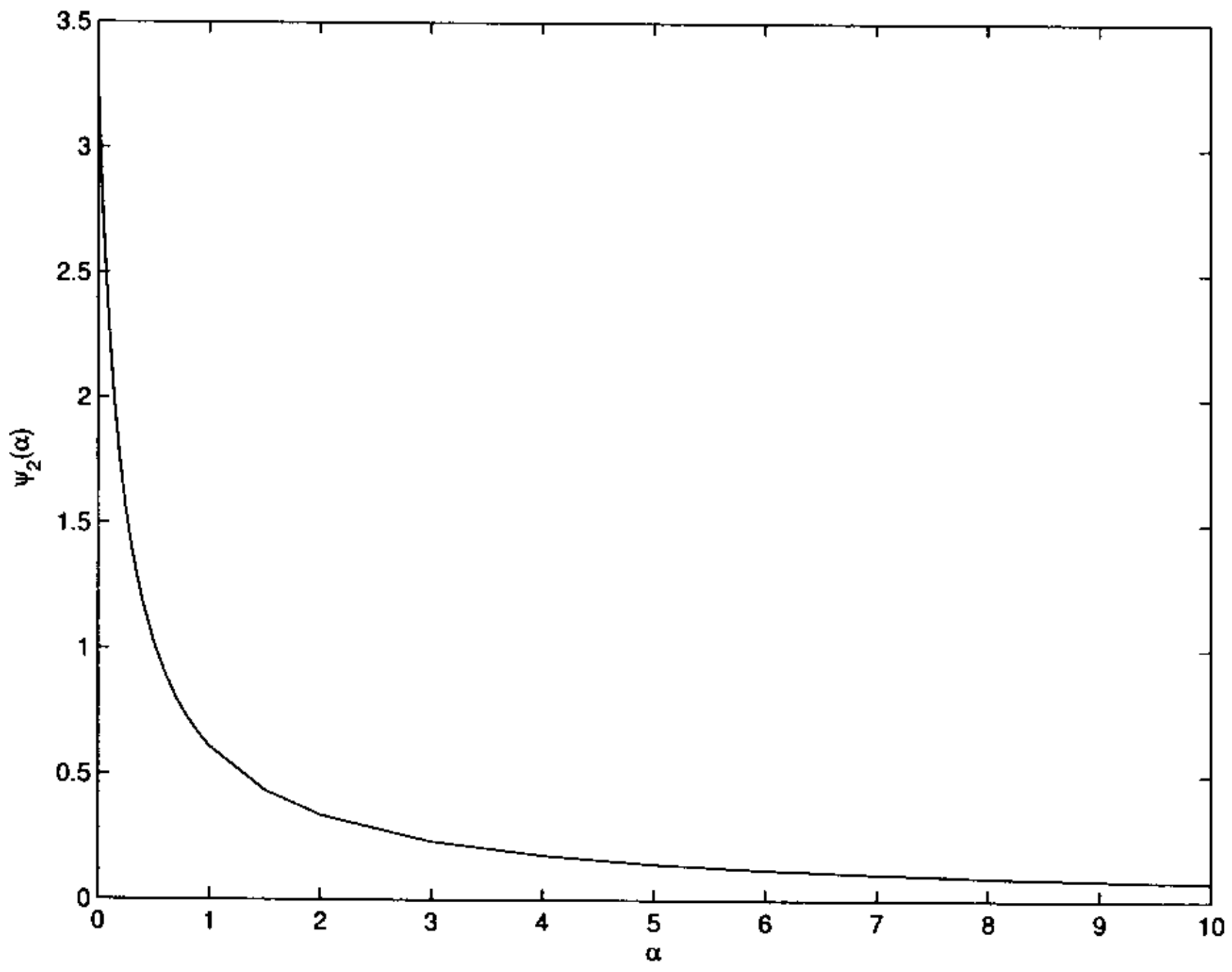

**Fig. 5:** Graph of function $\alpha \mapsto \psi_2(\alpha)$ in Eq. (26b).

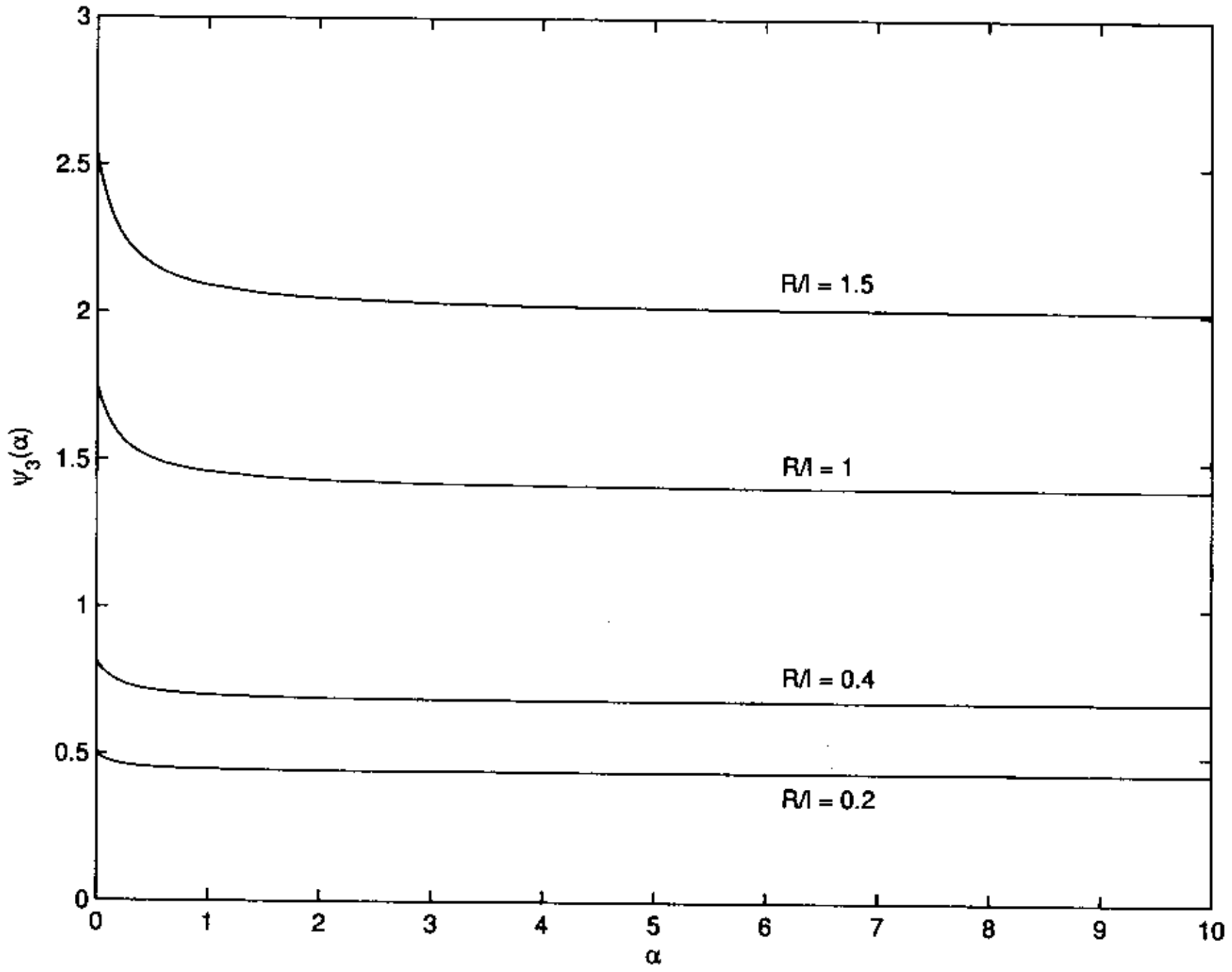

**Fig. 6:** Graph of function $\alpha \mapsto \psi_3(\alpha)$ in Eq. (26c).

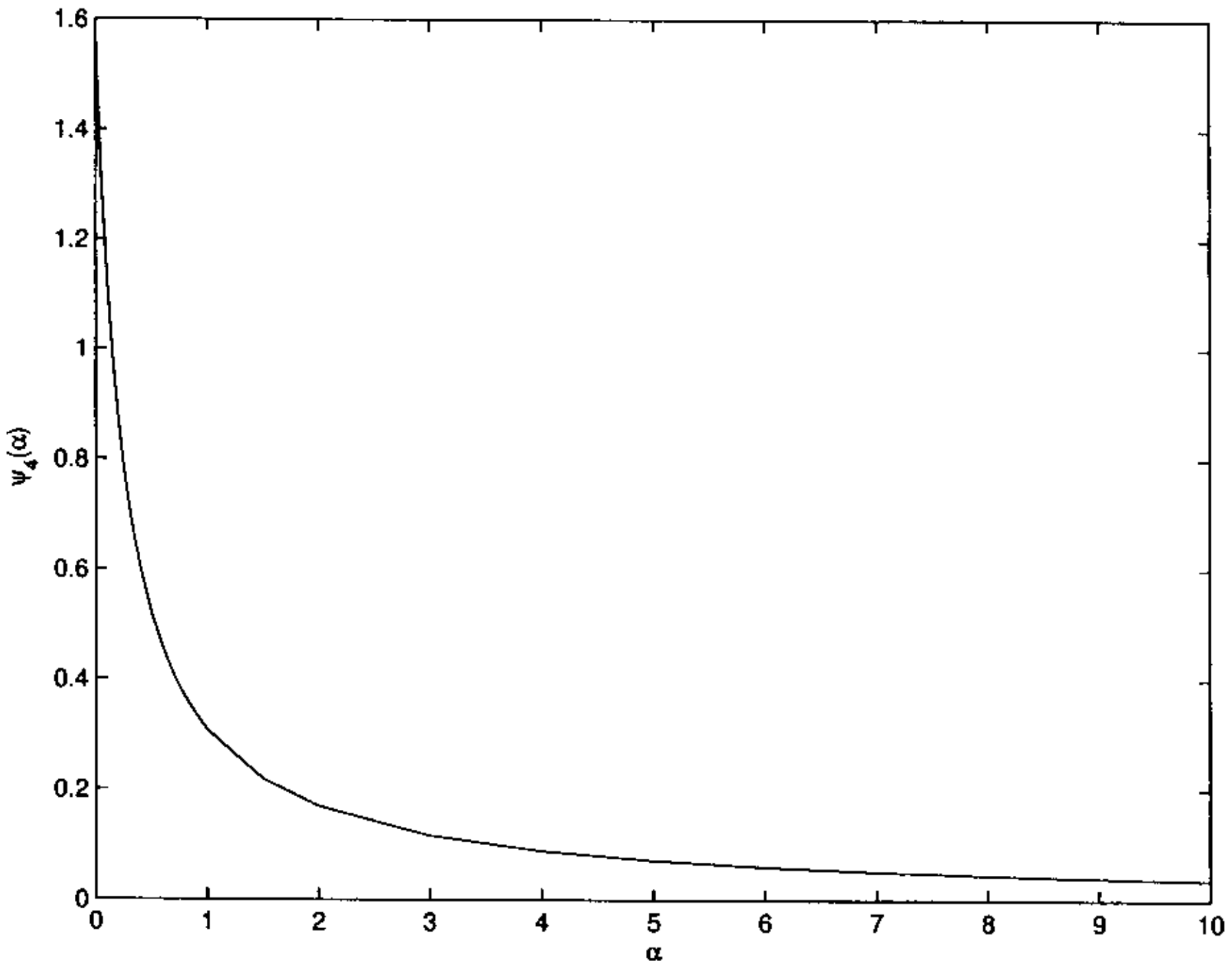

**Fig. 7:** Graph of function $\alpha \mapsto \psi_4(\alpha)$ in Eq. (26d).

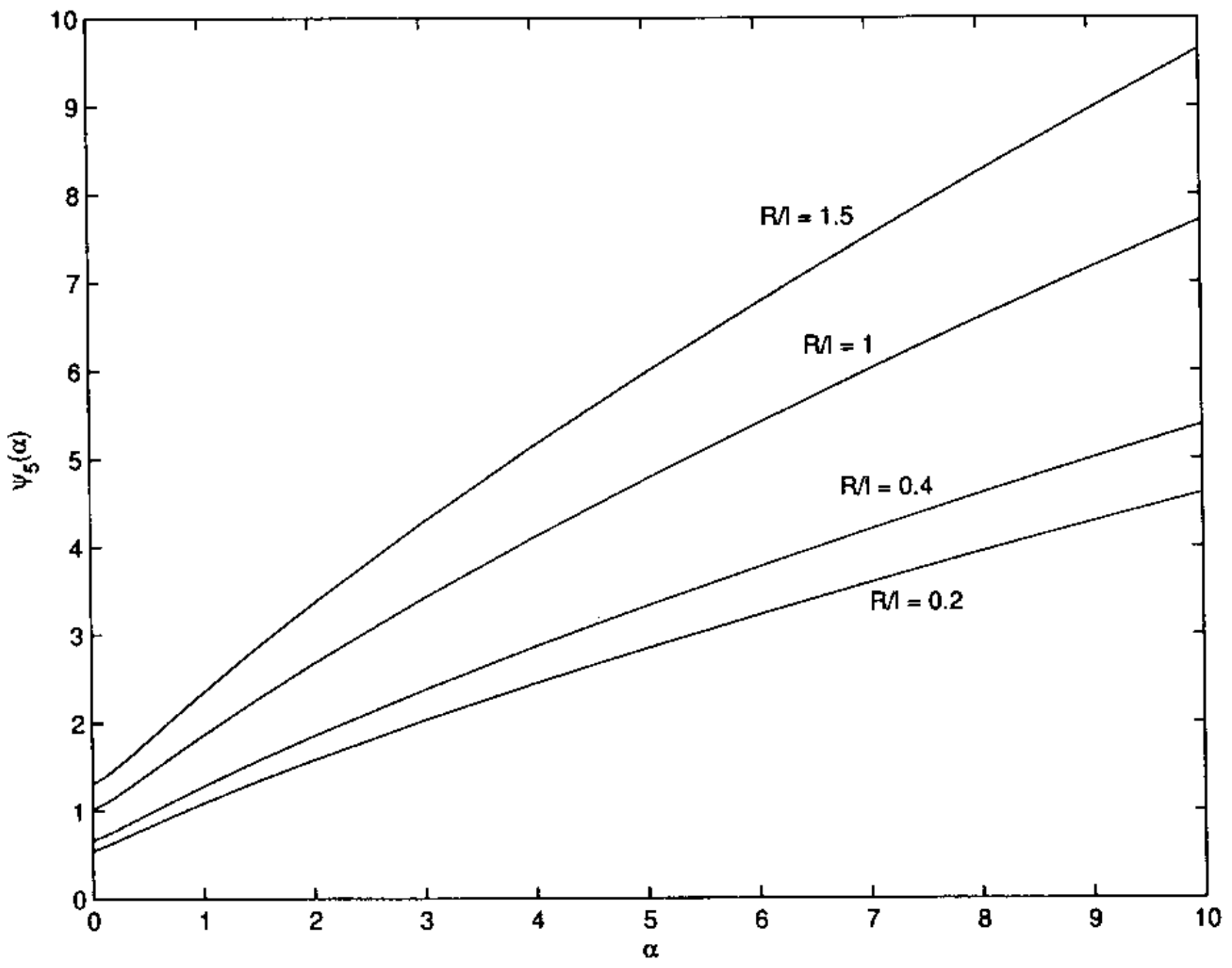

**Fig. 8:** Graph of function $\alpha \mapsto \psi_5(\alpha)$ in Eq. (26e).

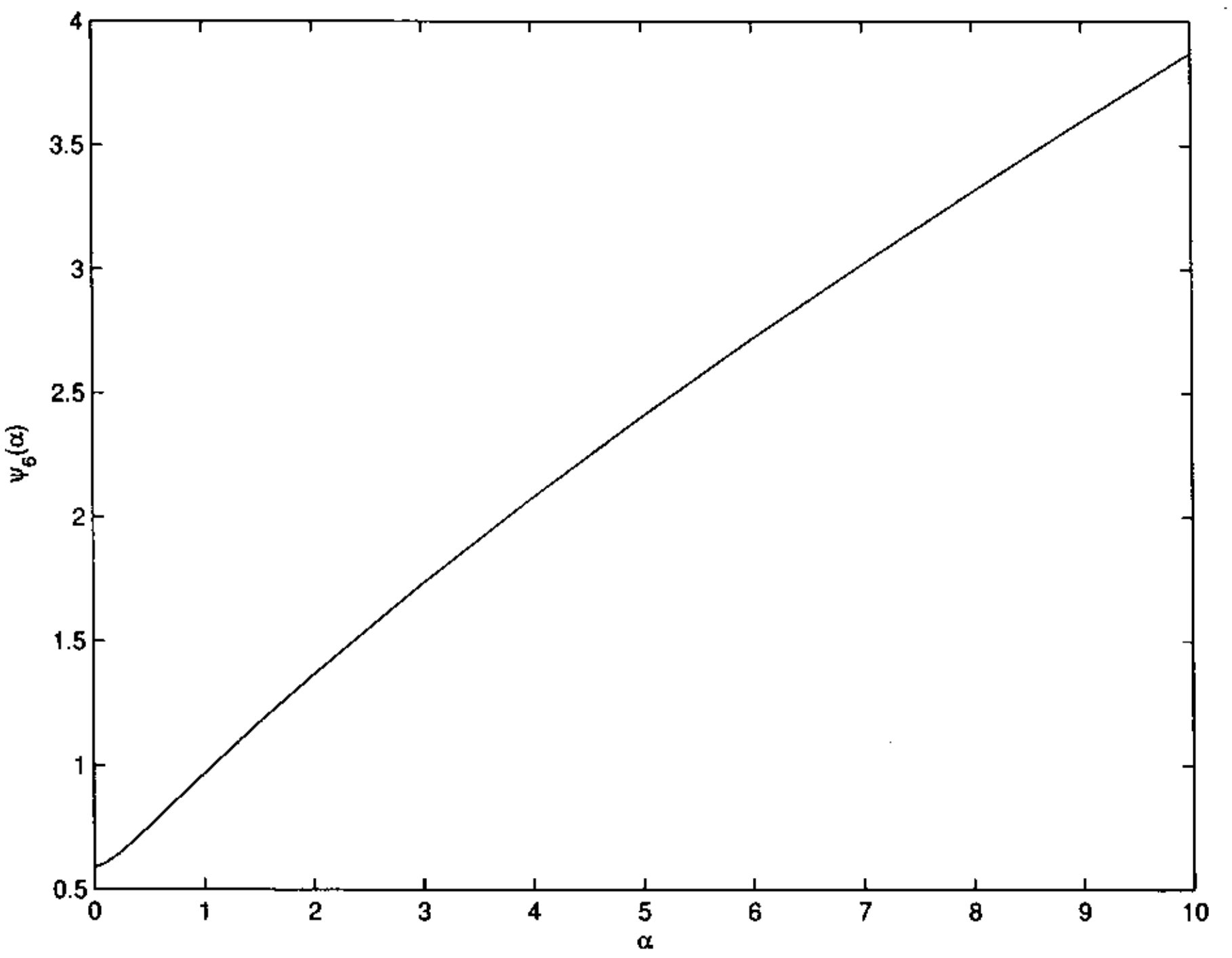

**Fig. 9:** Graph of function $\alpha \mapsto \psi_6(\alpha)$ in Eq. (26f).

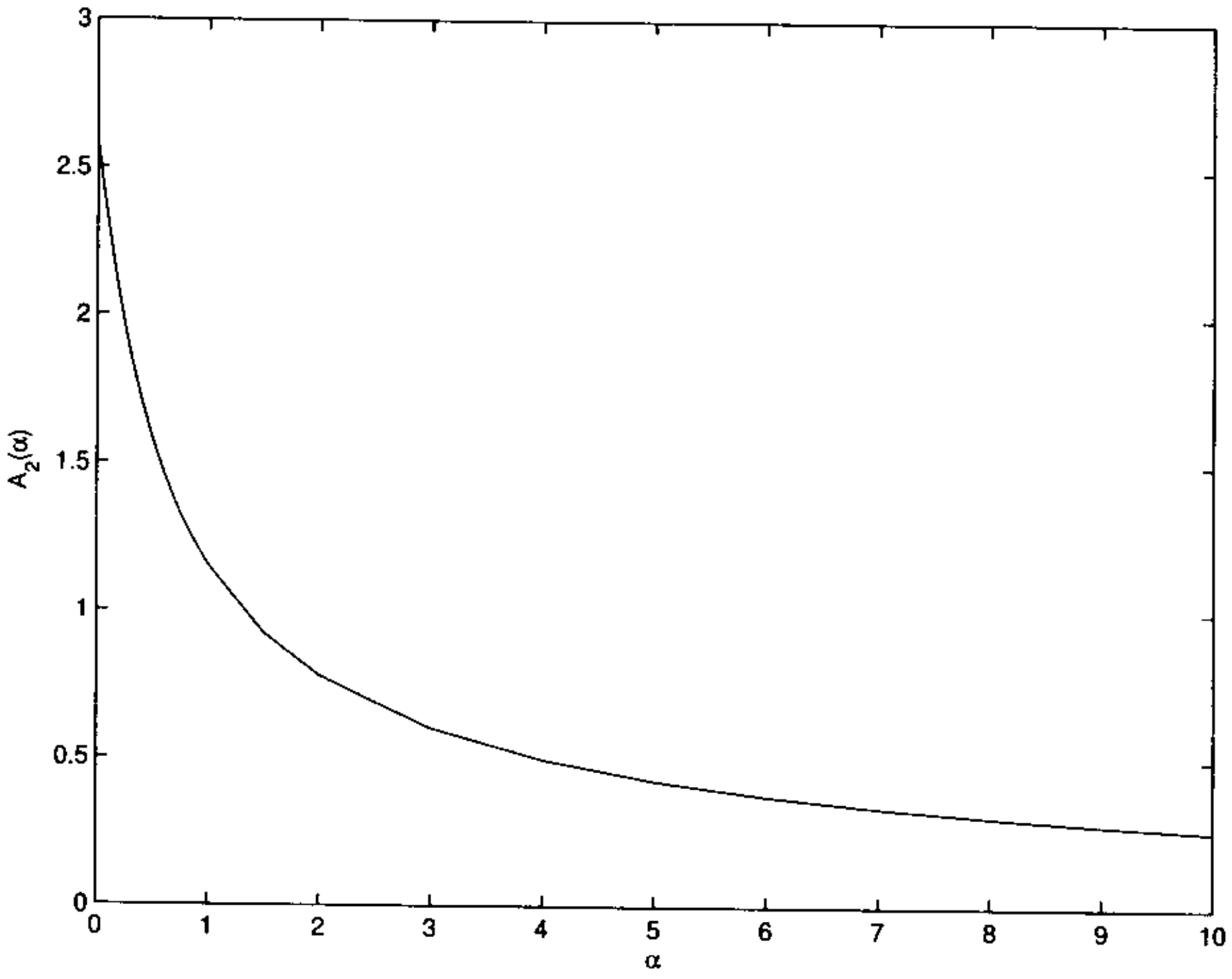

**Fig. 10:** Graph of $\alpha \mapsto A_2(\alpha)$ in Eq. (18b).

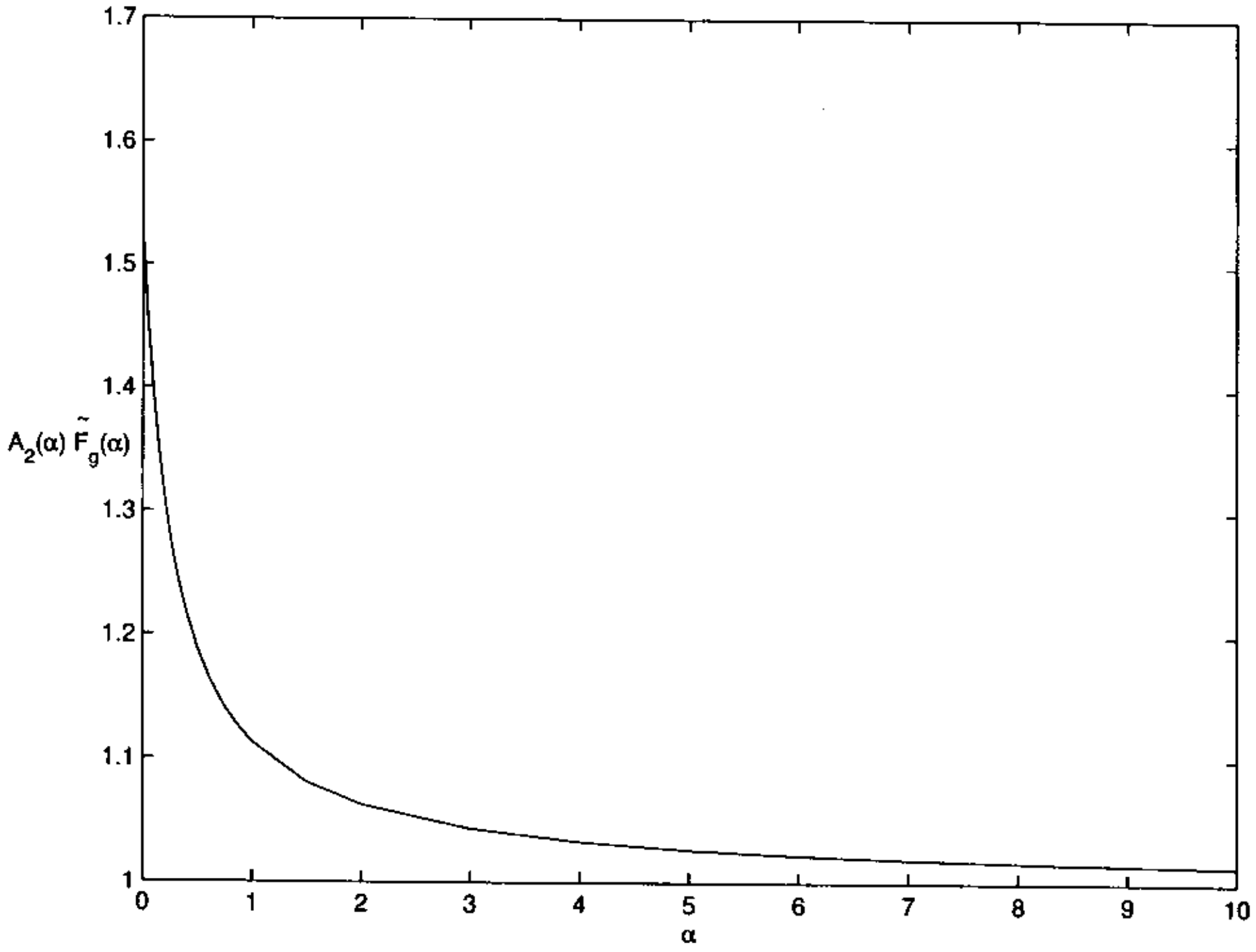

**Fig. 11:** Graph of function $\alpha \mapsto A_2(\alpha)\ \widetilde{F}_g(\alpha)$ .

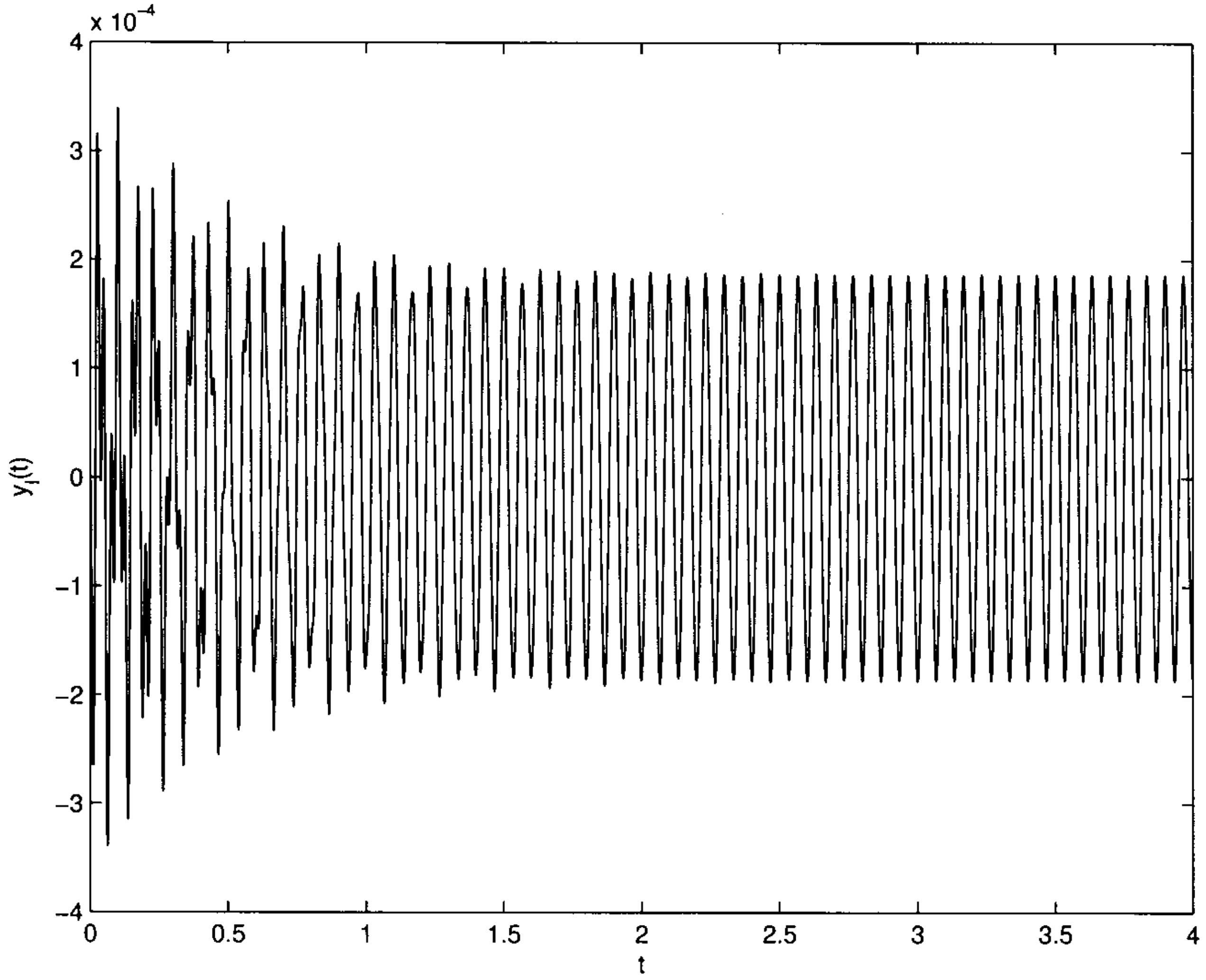

**Fig. 12:** The time history of the displacement of the tip of the beam in Example 4.1.

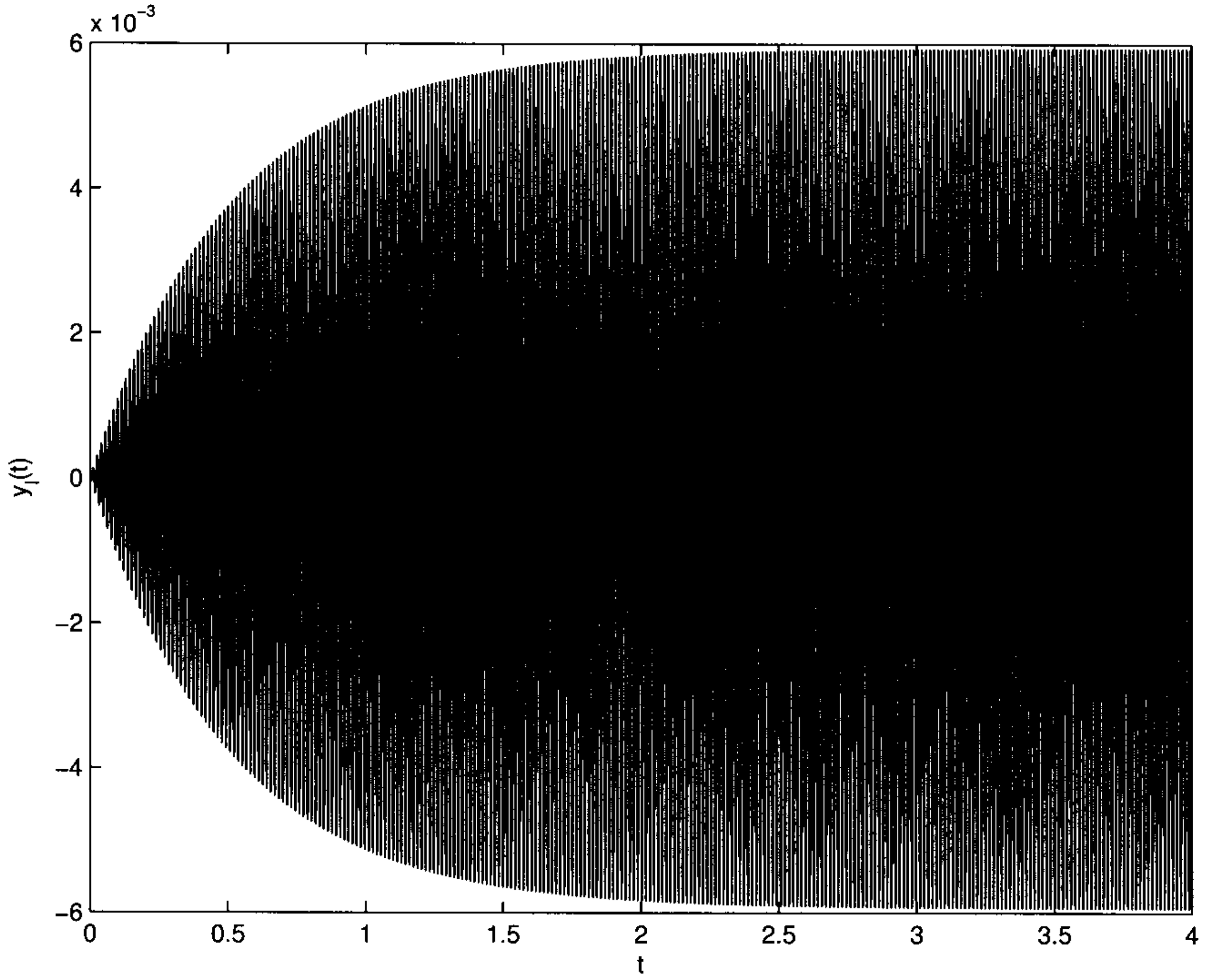

**Fig. 13:** The time history of the displacement of the tip of the beam in Example 4.2.

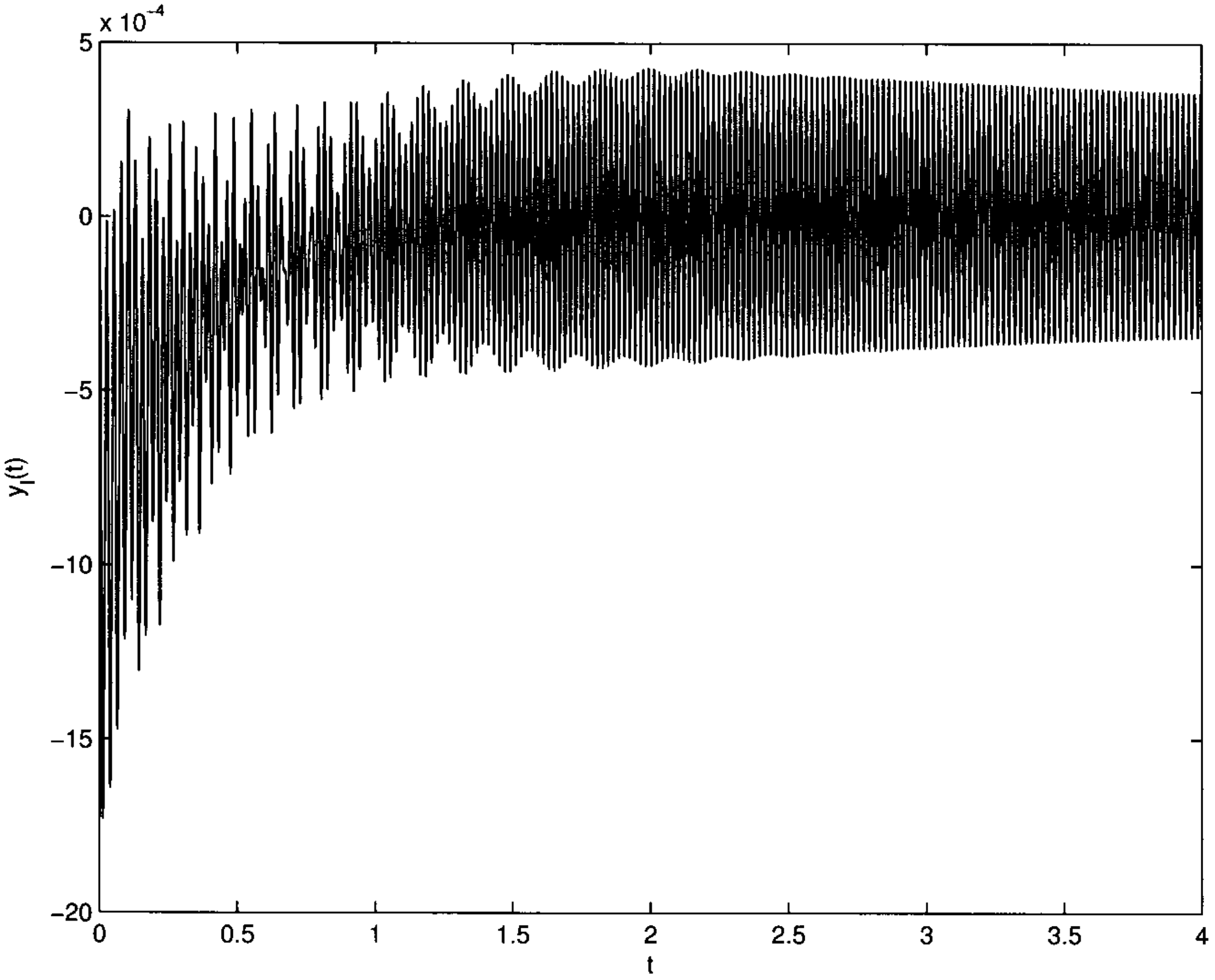

**Fig. 14:** The time history of the displacement of the tip of the beam in Example 4.3. The displacement is a harmonic function of time after a transient.